# A Constrained Tectonics Model for Coronal Heating

C. S. NG AND A. BHATTACHARJEE

Center for Magnetic Self-Organization

Center for Integrated Computation and Analysis of Reconnection and Turbulence

Institute for the Study of Earth, Ocean and Space

University of New Hampshire

Durham, New Hampshire, NH 03824

ABSTRACT

An analytical and numerical treatment is given of a constrained version of the tectonics model developed by Priest, Heyvaerts, & Title [2002]. We begin with an initial uniform magnetic field $\mathbf{B} = B_0\hat{\mathbf{z}}$ that is line-tied at the surfaces $z = 0$ and $z = L$. This initial configuration is twisted by photospheric footpoint motion that is assumed to depend on only one coordinate ($x$) transverse to the initial magnetic field. The geometric constraints imposed by our assumption precludes the occurrence of reconnection and secondary instabilities, but enables us to follow for long times the dissipation of energy due to the effects of resistivity and viscosity. In this limit, we demonstrate that when the coherence time of random photospheric footpoint motion is much smaller by several orders of magnitude compared with the resistive diffusion time, the heating due to Ohmic and viscous dissipation becomes independent of the resistivity of the plasma. Furthermore, we obtain scaling relations that suggest that even if reconnection and/or secondary instabilities were to limit the build-up of magnetic energy in such a model, the overall heating rate will still be independent of the resistivity.

*Subject headings*: Sun: corona --- MHD --- reconnection --- current sheets

1. INTRODUCTION

While a definitive resolution of the coronal heating problem continues to be elusive (see the recent review by Klimchuk [2006] and the monograph by Aschwanden [2005] for a comprehensive discussion), observations in recent years, especially from *Yohkoh*, *Solar and Heliospheric Observatory* (SOHO), and *Transition Region and Coronal Explorer* (TRACE) missions, have had a profound impact on our thinking regarding coronal heating mechanisms. The magnetic carpet, which covers the entire surface of the Sun and is constituted of magnetic fragments that are in a continual dynamical state of emergence, break-up, merging, and cancellation [Schrijver et al. 1998; Title 2000, Hagenaar 2001; Parnell 2001; Priest, Heyvaerts, & Title 2002], holds a key to understanding the heating of the global corona as well as the solar wind. Approximately 90% of the magnetic flux of the quiet Sun in the network concentrations, embedded in the carpet, originates from newly emerged bipolar pairs called ephemeral regions [Martin 1988] which have a mean total value of about $10^{19}$ Maxwells. The striking images produced by TRACE appear to suggest that the corona is composed of myriads of loops of various sizes, from large to small, with footpoints rooted in the network where most of the photospheric magnetic flux resides. It is estimated that 95% of the photospheric magnetic flux closes within the magnetic carpet (or the transition region) in low-lying loops, leaving only 5% to form large-scale connections [Schrijver & Zwaan 2000].

By measuring the rate of emergence of ephemeral regions from the *Michelson Doppler Imager* (MDI) instrument on board SOHO, it has been found that the photospheric flux in the quiet Sun is replaced approximately every 14 hours [Hagenaar 2001]. Surprisingly, however, the recycling time for magnetic flux in the solar corona is found to be only about 1.4 hours, which is about a tenth of the photospheric recycling time [Close et al. 2004]. This recycling time is obtained by considering the effects of reconnection as well as the emergence and cancellation of flux (which also involve a substantial amount of reconnection). These observations suggest a far more dynamic quiet-Sun corona than previously thought, with reconnection playing a crucial role in processing the flux and releasing magnetic free energy that may heat the global corona.

Fast reconnection is mediated by the formation of thin and intense current sheets. Parker [1972, 1994] proposed that the current density in the corona is distributed generically in the form of current sheets (tangential discontinuities). He has argued that the magnetic free energy of the system will be dissipated at near-Alfvenic rates at the sites of current sheets in the presence of a very small but finite resistivity, and has attempted to demonstrate that "…the X-ray luminosity of the Sun….is a consequence of a sea of small reconnection events ---nanoflares---in the local surfaces of tangential discontinuity throughout the bipolar magnetic fields of active regions" [Parker 1994]. Recently, Priest, Heyvaerts, & Title [2002] have proposed an analytical model in which a hierarchy of current sheets is formed at coronal separatrix surfaces, produced by the motions of a myriad of independent but small photospheric flux elements. By analogy with geophysical plate tectonics, where the relative motion of plates under the surface of the Earth produces potentially singular dynamics above the surface, Priest, Heyvaerts, &

Title [2002] have described their picture a "tectonics" model, a terminology we adopt in this paper.

Nearly all DC models of coronal heating are impulsive. This necessarily implies that in generic "nanoflare" events, large or small, there is a long time scale over which the magnetic energy is built up and stored (~ 1 day), followed by a much shorter time scale over which energy is dissipated. The latter may occur due to a combination of reconnection and secondary instabilities, which may involve consideration of collisionless mechanisms in the high-Lundquist-number corona (see Bhattacharjee 2004 for a review). The description of such complex coronal dynamics in three dimensions with adequate spatial and temporal resolution that can resolve physically relevant coronal plasma regimes is beyond the scope of present-day computers. However, *ab initio* global simulations within the resistive MHD framework with realistic boundary conditions have been carried out recently [Gudiksen & Nordlund, 2002, 2005a, 2005b]. Despite promising results from these global simulations, questions remain regarding the precise scaling of energy dissipation with respect to the resistivity of the plasma as well as the time-variability of the dissipation process. It is difficult to settle these questions within the scope of the global simulations because the dissipation in these simulations is not known with precision and controlled essentially by the level of numerical resolution. It thus appears that there is a need for complementary analytical and computational studies of coronal heating models that may be less complete than *ab initio* global simulations but enable the issues of scaling to be addressed with greater precision.

The main goal of this paper is to present an analytical and numerical treatment of a simple version of the tectonics model. We assume that closed, low-lying coronal loops which are anchored in the photosphere can be modeled in straight rectangular geometry. In other words, we neglect the curvature of magnetic loops on the Sun, and begin from an initial uniform magnetic field $\mathbf{B} = B_0 \hat{\mathbf{z}}$ that is line-tied at the surfaces $z = 0$ and $z = L$. This initial configuration is then twisted by photospheric footpoint motion that is assumed to depend on only one coordinate ($x$) transverse to the initial magnetic field. This strong assumption has the consequence that it enables us to describe the entire dynamics by a simple set of differential equations which are easily amenable to analytical and numerical solutions for prescribed footpoint motions. The geometric constraints imposed by our assumption precludes the occurrence of reconnection and secondary instabilities, but enables us to follow for long times the dissipation of energy due to the effects of resistivity and viscosity. In this limit, we delineate conditions under which the heating becomes independent of the resistivity of the plasma. Furthermore, we obtain scaling relations that suggest that even if reconnection and secondary instabilities were to limit the build-up of magnetic energy in such a model, the overall heating rate will still be independent of the resistivity.

The following is a layout of our paper. In Section 2, we describe the basic tectonics model. In Section 3, we obtain some exact analytical results in simple limits that are illustrative and provide useful benchmarks for our numerical simulations. In Section 4, we discuss our numerical results for random footpoint motions. We will provide heuristic estimates that explain our numerical results, and examine their implications for the coronal heating problem. We conclude in Section 4 with a summary, and a discussion of the proposed extensions of the present model.

## 2. CONSTRAINED TECTONICS MODEL

We assume that the coronal plasma is sufficiently low-beta that the effects of plasma pressure can be neglected, and that the dynamics can be described by the reduced MHD (RMHD) equations. The RMHD equations in dimensionless form can be written as

$$\frac{\partial \Omega}{\partial t} + [\phi, \Omega] = \frac{\partial J}{\partial z} + [A, J] + \nu \nabla_\perp^2 \Omega , \qquad (1)$$

$$\frac{\partial A}{\partial t} + [\phi, A] = \frac{\partial \phi}{\partial z} + \eta \nabla_\perp^2 A , \qquad (2)$$

where $\mathbf{B} = \hat{\mathbf{z}} + \mathbf{B}_\perp = \hat{\mathbf{z}} + \nabla_\perp A \times \hat{\mathbf{z}}$ is the magnetic field, $\mathbf{v} = \nabla_\perp \phi \times \hat{\mathbf{z}}$ is the fluid velocity, $\Omega = -\nabla_\perp^2 \phi$ is the $z$-component of the vorticity, $J = -\nabla_\perp^2 A$ is the $z$-component of the current density, $[\phi, A] \equiv \phi_y A_x - \phi_x A_y$, $\eta$ is the resistivity (inverse of the Lundquist number), and $\nu$ is the viscosity (inverse of the Reynolds number based on the Alfvén speed). The normalization adopted in equations (1) and (2) is such that magnetic field is in the unit of $B_z$ (assumed to be a constant in RMHD), velocity is the unit of $v_A = B_z/\sqrt{4\pi\rho}$ where $\rho$ is a constant density, length is in the unit of the transverse length scale $l$, the unit of time is $l/v_A$, $\eta$ is in the unit of $4\pi v_A l/c^2$, and $\nu$ is in the unit of $\rho v_A l$. An ideal magnetostatic equilibrium solution of equations (1) and (2) is obtained by setting all explicitly time-dependent terms as well as $\phi$ and $\eta$ to zero. We then obtain

$$\partial J / \partial z + [A, J] = 0 , \qquad (3)$$

which can also be written as $\mathbf{B} \cdot \nabla J = 0$. Equation (3) implies that the current density $J$ must be constant along a given magnetic field-line in an ideal static equilibrium.

Ordinarily, the problem of calculating time-dependent solutions of equations (1) and (2) in line-tied magnetic field geometry involves all three spatial coordinates and time. As a first step, we make the strong assumption that in addition to time $t$ and the coordinate *z* along which the magnetic field is line-tied, the dynamics depends on only one transverse coordinate *x*. Then the RMHD equations (1) and (2) reduce further to

$$\frac{\partial \phi}{\partial t} = \frac{\partial A}{\partial z} + \nu \frac{\partial^2 \phi}{\partial x^2}, \tag{4}$$

$$\frac{\partial A}{\partial t} = \frac{\partial \phi}{\partial z} + \eta \frac{\partial^2 A}{\partial x^2}. \tag{5}$$

Note that under this assumption, equations (4) and (5) become manifestly linear in $\phi$ and *A*. We assume that the system lies in a periodic box of width unity so that a field variable, say, *A* has the form

$$A(x,z,t) = \sum_{n=-\infty}^{\infty} A_n(z,t) e^{i2n\pi x}. \tag{6}$$

In this geometrically constrained model, the magnetic field transverse to *z* and the flow velocity are in the *y* direction only. Equations (4) and (5) will be solved subject to line-tied boundary conditions $\phi_n(0,t) = 0$ at $z = 0$, and $\phi_n(L,t) \equiv \phi_{nL}(t)$ at $z = L$. The photospheric motion described by the stream function $\phi_{nL}(t)$ is responsible for the build-up of magnetic energy in this system. Throughout this paper, we will use a boundary flow $v_y(x,L,t)$ with a step function-like shape as prescribed in the tectonics model to generate current sheets. This is important in the nonlinear dynamics of the model, but not as essential in our constrained model which does not have nonlinear terms.

We have developed a computer simulation code that integrates equations (4) and (5) numerically for arbitrary footpoint displacements. We use spectral decomposition in *x* and a leapfrog finite difference method in *z*. We use an implicit method for time-integration so that we can take larger time steps than is allowed by the Courant-Friedrich-Lewey condition for numerical stability of explicit methods. This enables us to integrate efficiently for long periods of time in order to obtain good statistics. Since equations (4) and (5) are linear and do not permit nonlinear energy transfer and cascade due to mode-coupling, the resolution we adopt initially works well for the entire duration of the simulation. This also means that heating mechanisms that rely on a current cascade mechanism (van Ballegooijen 1986, Hendrix et al. 1996, Galsgaard and Nordlund 1996) cannot be described within the constraints of the present model.

## 3. CONSTANT FOOTPOINT DRIVE: A SIMPLE EXAMPLE

As discussed in Section 1, we are observationally motivated to consider photospheric footpoint motions specified by the function $\phi_{nL}(t)$ with a coherence time $\tau_{coh}$ of the order of hours, which is much less than the resistive diffusion time $\tau_r = w^2/\eta$, where $w$ is the characteristic length scale of photospheric motion. Thus, $\phi_{nL}(t)$ cannot be regarded as constant in time. However, the case of constant footpoint drive $\phi_{nL}$ is analytically tractable, illustrative, and useful for benchmarking our numerical solutions. For such a constant drive, one would expect that the system would build up to an asymptotic steady state whereby the energy injected into the plasma by footpoint motion will be balanced by dissipation. In the absence of reconnection and/or instabilities, this balance is typically realized on the time scale of resistive diffusion. Setting $\partial/\partial t \to 0$ in equations (4) and (5), we obtain the exact analytic solutions

$$\phi_n(z) = \phi_{nL} \frac{\sinh\left(\sqrt{\eta v}\, k_n^2 z\right)}{\sinh\left(\sqrt{\eta v}\, k_n^2 L\right)} , \qquad (7)$$

$$A_n(z) = \phi_{nL} \sqrt{\frac{v}{\eta}} \frac{\cosh\left(\sqrt{\eta v}\, k_n^2 z\right)}{\sinh\left(\sqrt{\eta v}\, k_n^2 L\right)} , \qquad (8)$$

where $k_n = 2n\pi$. For very small resistivity, if we assume that $\sqrt{\eta\nu}k_n^2 L \ll 1$, equation (8) yields the result

$$A_n(z) \xrightarrow{\eta \to 0} \frac{\phi_{nL}}{\eta k_n^2 L}, \tag{9}$$

which is independent of z, and hence describes an equilibrium state that obeys equation (3). Note that the saturated magnetic field does not depend on viscosity, but depends inversely on resistivity.

Equation (9) suggests that the transverse magnetic field in steady state obeys the scaling relation

$$\frac{B_\perp}{B_z} \sim \frac{v_L \tau_r}{L} \equiv \frac{l_r}{L}, \tag{10}$$

where $v_L$ is the typical photospheric velocity and $l_r$ is the typical distance a photospheric footpoint moves in a resistive diffusion time, $\tau_r \sim w^2/\eta$, where $w$ is an average length scale. Strictly speaking, the length scale $w$ in equation (10) is dependent on the wave number $k_n$; here for the purpose of estimation we use an average length that is a fraction of the width of a field cell. The ratio in equation (10) is very large for very small values of resistivity. The Ohmic dissipation rate is

$$W_d = \eta \int J^2 d^3 x \to \frac{1}{\eta L} \sum_{n=-\infty}^{\infty} |\phi_{nL}|^2, \tag{11}$$

which is inversely proportional to resistivity and can also be very large. This implies that the energy injection rate per unit area in the photosphere is of the order of

$$I_p \sim \frac{W_d}{w^2} \sim B_z^2 v_L \frac{l_r}{L}. \tag{12}$$

Equation (11), although not physically realistic for the corona, is useful as a test example for benchmarking simulations as well as for providing further motivation for the studies described in Section 4.

We test our simulation code with $\phi_{nL}(t)$ constant, discussed above. The solution evolves and tends to the analytic form given by equations (7) and (8) when $t \geq \tau_r$, as expected. Figure 1(a) shows the boundary flow velocity $v_y(x,L)$, derived from $\phi_L(x)$, for a case with $\eta = 5 \times 10^{-6}$, $\nu = 10^{-5}$, $L = 10$, computed with a spatial grid $256 \times 128$ in $(x,z)$-space. The transverse magnetic field $B_y(x,L)$ and current density $J(x,L)$ in the steady state at $z = L$ are plotted in Figures 1(b) and 1(c), respectively. We see that even though $v_y(x,L)$ is nearly step function-like, $J(x,L)$ is globally quite smooth except in a localized region near the tips of the profile. As a result, $B_y(x,L)$, which is obtained by integrating $J(x,L)$, is smooth as well. The transverse projection of $B_y(x,z)$ for all $z$ has almost the same profile, that is, it is almost independent of $z$, as expected from equation (9). An estimate for the magnitude of the transverse magnetic field may be obtained by substituting $B_z = 1$, $v_L \approx 0.0018$, $\tau_r \sim w^2/\eta$ whence $w \sim 0.2$ (so that $\tau_r \sim 8000$ and $l_r \sim 14$) into equation (10), which yields $B_y \sim 1.4$, approximately consistent with the magnitude shown in Figure 1(b).

The $J$ or $B_y$ profiles are significantly smoother than the vorticity profile $\Omega(x,L)$, shown in Figure 1(d), which exhibits sharp sheet-like structures. Sheet-like structures in $J$ and $\Omega$ are strongest at the model photosphere, $z = L$, and become smoother as $z$ decreases slightly from $L$, as shown in Figures 1(e) and (f) for $z = 0.9L$. Contour plots of $J$ and $\Omega$ for the whole $x$-$z$ region is in Figures 1(g) and (h), with a rainbow color scale which uses black/dark purple for most negative contours and red for most positive contours. One can compare these to the plots in Figures 1(c) to (d) to obtain a more quantitative correspondence between colors and actual values. In what follows, we will use this color scheme for other case studies as well.

## 4. RANDOM FOOTPOINT DRIVE

The exercise in Section 3, although primarily of academic interest, underscores the need to consider a time-varying $\phi_{nL}(t)$ with magnitude and direction randomized in time, characterized by a coherence time scale, $\tau_{coh}$. Indeed, observations of the magnetic carpet cited in Section 1 suggest that the shuffling of photospheric footpoints is a near-random process, where velocities at the boundaries of supergranular cells appear to have large changes in spatial derivative. Hence, we will consider a simple model of photospheric motion in which the velocity field has step function-like behavior at the boundaries of cells. We write $\phi_L(x,t)$ in the form

$$\phi_L(x,t) = \phi_0(t) \sum_{n=1}^{N} \frac{(-1)^n}{(2n-1)^2} \sin[(2n-1)2\pi x], \qquad (13)$$

where $N$ is a large number, and a time series $\phi_0(t) = \overline{\phi}_0 \cos[\theta(t)]$ is produced by a random walk process,

$$\theta(t + \Delta t) = \theta(t) + \pi \sqrt{\Delta t / \tau_{coh}} \, \text{rand}(-1,1). \qquad (14)$$

We choose $\overline{\phi}_0$ to be a constant small enough that the photospheric velocity derived from equation (13) is much less than the Alfvén speed ($v_L << v_A$). The random function rand(-1,1) returns a uniformly distributed random number between -1 and 1, and $\Delta t$ is chosen to be much smaller than $\tau_{coh}$. Note that the case of constant footpoint drive in time, discussed in Section 3, is a special case of equation (14) with $\tau_{coh} \to \infty$ and $\theta(0) = 0$. We have also used different random series generators, similar to those used by Longcope & Sudan [1994], and showed that our results do not depend on how the random series is generated.

We will simulate a case with $\tau_{coh} = 1000$, which is about an order of magnitude smaller than $\tau_r$, with other parameters the same as in Section 3. This value of $\tau_{coh}$ is of the order of $w/v_L$ used in the simulation. While $\tau_r$ is larger than $\tau_{coh}$, it is still much smaller than the typical resistive diffusion time under realistic coronal conditions for which $\tau_r$ can be very large ($\sim 10^{12}$-$10^{13}$) if we use estimates for classical Spitzer resistivity. Since $\phi_L(x,t)$ depends on time explicitly, the solution will not strictly attain a steady state. However, after about a resistive diffusion time, the solution fluctuates around a certain average level that can be regarded as a statistical steady state. Figures 2(a)-(h) show plots of the same physical quantities in a statistical steady state as those plotted in Figures 1(a)-(h). Figure 2(a) shows a snapshot in time of the photospheric driving velocity, which is evidently quite different than that shown in Figure 1(a), both in magnitude and direction. Figures 2(b) and (c) show that the statistical levels of $B_y$ and $J$ are only fractions of the saturated levels shown in Figures 1(b) and (c). The current sheet-like structure at the tips of the $J(x,L)$ profile, seen in Figure 2(c), is more conspicuous than that in Figure 1(c). However, just as in Figure 1(e), the sharpness of the tips in Figure 2(e) is reduced when $z$ is slightly off the boundary. From the contour plot of $J(x,z)$ in Figure 1(g), we note that it is still approximately independent of $z$, and thus describes a quasi-equilibrium state.

The average energy dissipation rate can be defined as

$$\overline{W}_d(t) \equiv \frac{1}{t}\int_0^t W_d(t')dt' = \frac{1}{t}\int_0^t \int \left[\eta J^2(\mathbf{x},t') + \nu\Omega^2(\mathbf{x},t')\right]d^3x\,dt', \quad (15)$$

and is plotted in Figure 3 as the black trace. This level is over an order of magnitude smaller than that of the steady state depicted in Figure 1 for which $W_d = 0.0164$. To see how the energy dissipation rate scales with resistivity, we repeat the same run with all parameters fixed except the resistivity which is changed to $\eta = 10^{-5}$ (red trace), and then to $\eta = 2\times 10^{-5}$ (blue trace). We note that $\overline{W}_d$ decreases as $\eta$ increases, which appears to be qualitatively similar to the trend seen in the case of constant footpoint drive (which yields $W_d \propto 1/\eta$), except that the dependence of $\overline{W}_d$ on resistivity is much weaker in this case. Note that this result, which is an average power over many coherence times, $\tau_{coh}$, differs qualitatively from the analytical results of Priest, Heyvaerts, and Title [2002] who obtain $\overline{W}_d \propto \eta^{1/2}$ in the limit of small $\tau_{coh}$, averaged over only one $\tau_{coh}$ starting from an initial potential field.

In view of the results reported above, where the dependency of the energy dissipation rate on resistivity appears to be weakened by reducing the coherence time, we are motivated to consider the case in which $\tau_{coh}$ is much smaller than $\tau_r$. Therefore, we simulate a case with $\tau_{coh} = 20$. Figure 4 shows the time-evolution of $\overline{W}_d$. We note that the differences in $\overline{W}_d$ between the three cases with $\eta = 5\times 10^{-6}$, $10^{-5}$, and $2\times 10^{-5}$ are very small, suggesting that $\overline{W}_d$ tends to become approximately independent of resistivity in the limit of small $\tau_{coh}$, which is one of the main results of this paper. Note that the primary contribution to the average heating rate is due to the Ohmic dissipation, not viscous dissipation.

To develop a qualitative understanding of the weak dependence of $\overline{W}_d$ on $\eta$, we consider the following scaling estimates based on random walk statistics. Let us consider a time interval $t$ in which $N$ random steps are taken by a magnetic field line due to random photospheric footpoint motions at $z = L$ characterized by a small coherence time $\tau_{coh}$, such that $t = N\tau_{coh}$. The root-mean square transverse distance moved by a field line at $z = L$ under this random motion is given by

$$\ell_{rms} = \sqrt{N} v_L \tau_{coh} = v_L \sqrt{t \tau_{coh}}. \tag{16}$$

In a time scale of the order of the resistive diffusion time $\tau_r$, which is the time scale on which the system attains a statistical steady state, the root-mean-square transverse distance is given by $\ell_{rms} \sim v_L \sqrt{\tau_r \tau_{coh}}$. Then, the average perpendicular magnetic field strength is estimated to be

$$\frac{\overline{B}_y}{B_z} \sim \frac{\ell_{rms}}{L} \sim \frac{v_L}{L} \sqrt{\frac{\tau_{coh} w^2}{\eta}}. \tag{17}$$

Since the current density $J$ is still mostly quite smooth over the whole field cell, we can estimate $\overline{W}_d$ by using $\overline{J} \sim \overline{B}_y / w$, which yields,

$$\overline{W}_d \sim \eta \int \overline{J}^2 d^3x \sim \eta(\overline{B}_y^2 / w^2)(Lw^2) \sim \frac{v_L^2}{L} B_z^2 \tau_{coh} w^2. \tag{18}$$

Note that the right side of expression (18) is manifestly independent of $\eta$, as suggested by the numerical result shown in Figure 4. Note also that this average heating rate is not the same as the energy transferred from the instantaneous Poynting flux, which is proportional to $B_y$, and thus $\eta^{-1/2}$. This is because the system is in a statistical steady state, not a true steady state. So, instantaneous Poynting flux does not have to balance the heating rate. Instead most of the flux contributes to changing (increasing or decreasing) the total magnetic energy of the system, which is actually proportional to $\eta^{-1}$. Thus, the relative change due to the instantaneous Poynting flux is still small. In fact, the instantaneous Poynting flux can be positive or negative. It is the time-averaged value that will turn out to be positive and will balance the average heating rate.

From equation (17), we infer that $\eta^{1/2}\overline{B}_y$ should be approximately constant, if the parameters on the right of equation (17) are held fixed. This is indeed consistent with the numerical results presented in Figure 5. If we substitute the simulation parameters, that is, $B_z = 1$, $L = 10$, $v_L \approx 0.0018$, $\tau_{coh} = 20$, and $w \sim 0.2$ into equation (17), we recover approximately the asymptotic value plotted in Figure 5 for the quantity $\eta^{1/2}\overline{B}_y$. A similar exercise with the average energy dissipation rate $\overline{W}_d$, predicted by equation (18), yields a number within a factor of two of the asymptotic dissipation rate seen in Figure 4.

We note that since $\overline{B}_y$ is inversely proportional to $\eta^{1/2}$, it can get very large compare with $B_z$ if we assume that the resistivity of the corona is given by the classical Spitzer resistivity which is very small, say, $\eta \sim 10^{-13}$. This is not evident in the runs reported in this paper so far because we have not used values of the resistivity that are small enough. To be specific, we have obtained $\overline{B}_y \sim 0.087, 0.062, 0.044$ for the three levels of resistivity in the runs with $\tau_{coh} = 20$ reported in Figure 5. To see what would happen for very small $\eta$, we also run a case with $\eta = 10^{-10}$ and $\tau_{coh} = 600$, with other parameters the same as those in Figure 5. Since $\tau_r$ is so large in this case, we cannot run this case long enough to realize a statistical steady state. Nonetheless, Figures 6 (a) and (b), which show time series for $\overline{W}_d$ and $\overline{B}_y$ for this run, point towards the trend. We see that at very small values of the resistivity, the quantities $\overline{W}_d$ and $\overline{B}_y$ fluctuate but generally increase with time with no apparent saturation. While $\overline{W}_d$ has not increased to the same level as shown in Figure 4 for the duration of the simulation, $\overline{B}_y$ has increased to a value more than twenty times larger than $B_z$.

We expect that the realization of such large values of $\overline{B}_y$ will be thwarted by the intervention of reconnection and/or instabilities in a 3D dynamical calculation. However, although the corona is not expected to reach the statistically stationary state with such a large $\overline{B}_y$, it is interesting to note that the heating power delivered per unit area of the solar coronal surface is not sensitive to the threshold at which such activity intervenes. Let us assume that the build-up of $\overline{B}_y$ occurs until the relation $\overline{B}_y = fB_z$ is satisfied in a time $t \sim \tau_E$, where $f$ is a dimensionless number. Then the magnetic energy built up during this process will be quickly dissipated in a time much shorter than $\tau_E$ due to reconnection and/or instabilities. Using the same type of scaling arguments based on random footpoint statistics that led to equation (17), we obtain

$$f \equiv \frac{\overline{B}_y}{B_z} \sim \frac{v_L}{L}\sqrt{\tau_{coh}\tau_E} \; , \tag{19}$$

or

$$\tau_E \sim \left(\frac{fL}{v_L}\right)^2 \frac{1}{\tau_{coh}} \; . \tag{20}$$

Therefore, the average energy dissipation rate $\overline{W}_d$ is given by

$$\overline{W}_d \sim \frac{1}{\tau_E}\int \overline{B}_y^2 d^3x \sim \left(\frac{v_L}{fL}\right)^2 \tau_{coh}(fB_z)^2(Lw^2) \sim \frac{v_L^2}{L}B_z^2\tau_{coh}w^2 \; , \tag{21}$$

which is the same as that obtained in Eq. (18). Note that this average energy dissipation rate is independent of the threshold factor $f$, as well as the resistivity $\eta$. The heating rate per unit area is thus given by

$$I_p \sim \frac{\overline{W_d}}{w^2} \sim B_z^2 v_L \frac{v_L \tau_{coh}}{L} \quad . \tag{22}$$

Using the parameters given by Priest, Heyvaerts, and Title [2002], this heating rate can be sufficient to account for observations if the ratio $v_L \tau_{coh}/L$ is in the approximate range 0.25 - 0.5, which is physically plausible for a time-variable Sun, covering quiet as well as active regions. Note also that if we define an effective $B_{yeff} = B_z v_L \tau_{coh}/L$, the heating rate given by Eq. (22) is the same as an effective Poynting flux

$$I_p \sim B_z B_{yeff} v_L. \tag{23}$$

So in order to account for observed heating rate, we need to have $\tan^{-1}(B_{yeff}/B_z)$ in the range around $20°$, consistent with the estimation given by Klimchuk [2006].

5. SUMMARY

In this paper, we have developed a constrained version of the tectonics model of Priest, Heyvaerts, and Title [2002]. We call it a constrained model because it is constrained geometrically to be two-dimensional, where an initially constant magnetic field twisted by photospheric footpoint motion that is assumed to depend on only one coordinate transverse to the initial magnetic field. The geometric constraints imposed by this strong assumption precludes the occurrence of reconnection and secondary instabilities, but enables us to follow for long times the dissipation of energy due to the effects of resistivity and viscosity. In this limit, we demonstrate that when the coherence time of random photospheric footpoint motion is much smaller by several orders of magnitude compared with the resistive diffusion time, the heating due to Ohmic and viscous dissipation becomes independent of the resistivity of the plasma. Furthermore, we obtain scaling relations suggesting that even if reconnection and secondary instabilities were to limit the build-up of magnetic energy in such a model, the overall heating rate will still be independent of the resistivity.

Although our model is simple, it is encouraging as a first step in the development of a more complete 3D model that can produce quantitative information on the scaling of coronal heating with respect to the dissipation mechanism. Such a model, which will enable richer dynamics, will undoubtedly introduce new dynamical features not present in the present treatment. In a more complete model, we expect that reconnection and/or instabilities will intervene much before a resistive diffusion time scale. For instance, reconnection should play a strong role within a characteristic Sweet-Parker time scale in a resistive MHD model, and a statistical steady state, with external drive driving balanced by dissipation, should be attainable on time scales much shorter than the time scale of resistive diffusion. However, we conjecture that the scaling relations on the energy dissipation and average heating rate obtained in this paper will continue to hold approximately even in a more complete 3D model. From the perspective of the observations discussed in Section 1, we conjecture that current and vortex sheets, which mediate reconnection and/or instabilities at low altitudes in a tectonics model, appear to have the potential to account for the coronal heating budget. In future work, we will test this conjecture in a more complete dynamical model.

Acknowledgement

This research is supported by National Science Foundation Grant Nos. AST-0434322 and ATM-0422764, and the Department of Energy under the auspices of the Center for Magnetic Self-Organization (CMSO) and the Center for Integrated Computation and Analysis of Reconnection and Turbulence (CICART).

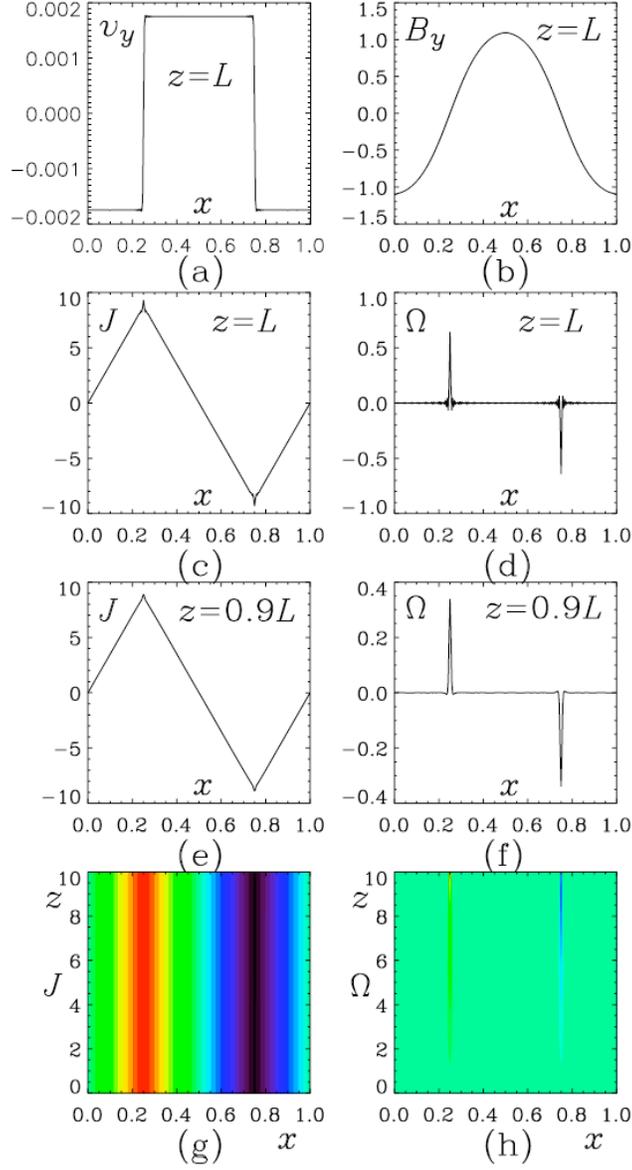

Figure 1. Saturated state for a run with constant footpoint drive for $\eta = 5 \times 10^{-6}$, $\nu = 10^{-5}$, $L = 10$. (a) Boundary flow velocity $v_y(x,L)$. (b) Transverse magnetic field $B_y(x,L)$. (c) Current density $J(x,L)$. (d) Vorticity $\Omega(x,L)$. (e) Current density $J(x,0.9L)$. (f) Vorticity $\Omega(x,0.9L)$. (g) Contour plot of the current density $J(x,z)$ using a rainbow color scale: black/dark purple for most negative contours and red for most positive contours. (h) Contour plot of the vorticity $\Omega(x,y)$.

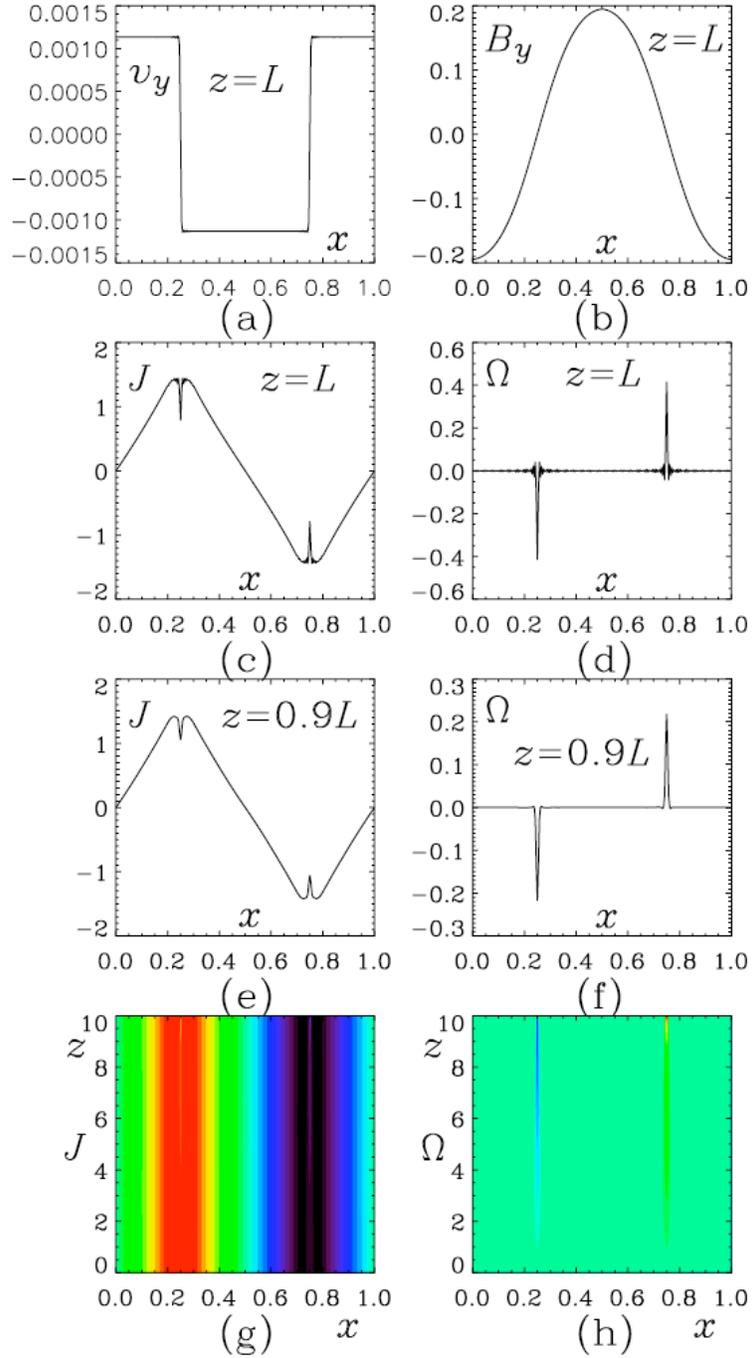

Figure 2. Plots of the same quantities as in Figure 1 for a run using random boundary flow with coherence time $\tau_{coh} = 1000$, at a time when the system has attained a statistical steady state. Other parameters are the same as the case shown in Figure 1.

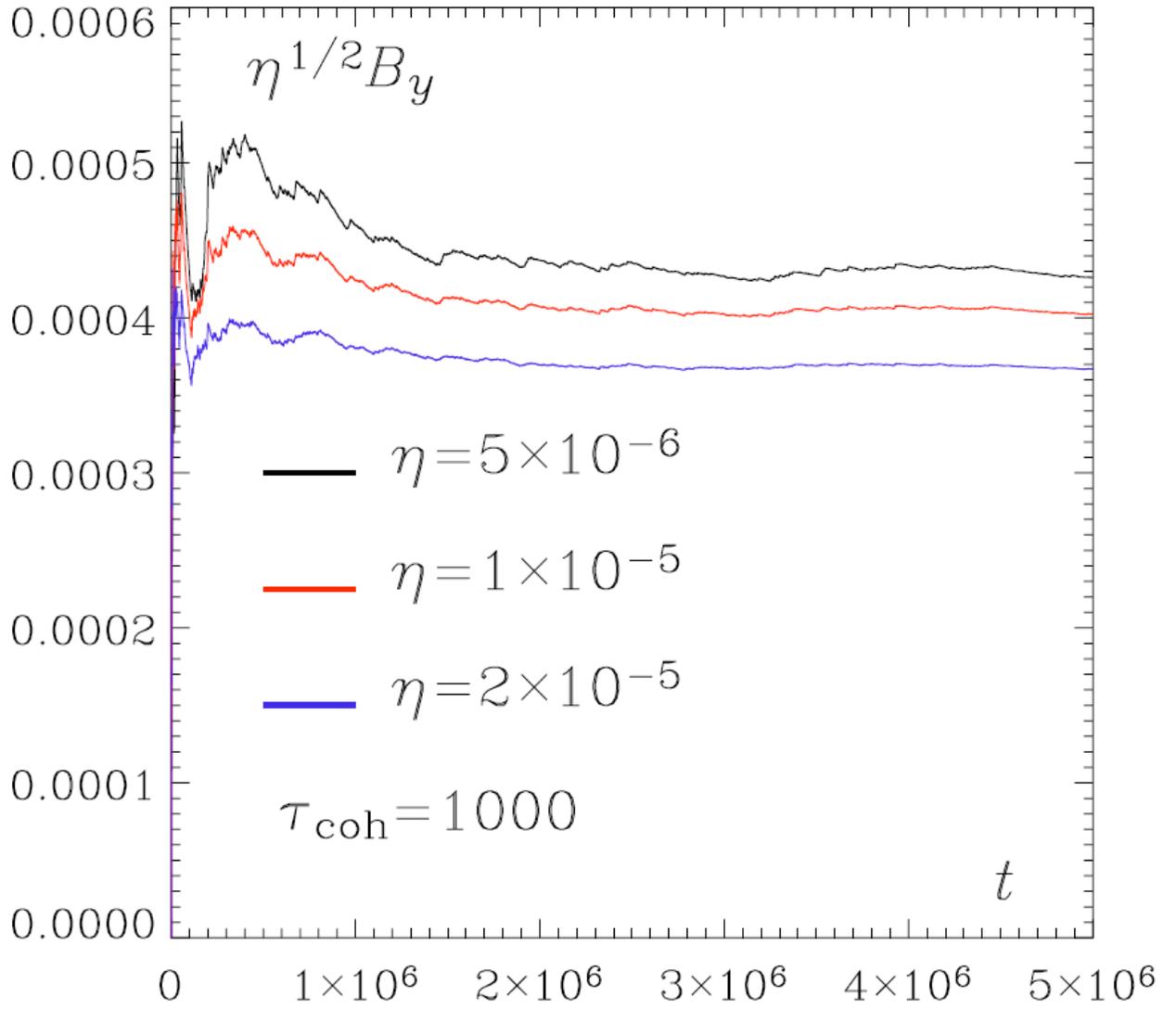

Figure 3. The time-averaged energy dissipation rate $\overline{W}_d$ as a function of time for the case shown in Figure 2 (black trace). Also plotted are $\overline{W}_d$ for $\eta = 10^{-5}$ (red trace), and $\eta = 2 \times 10^{-5}$ (blue trace), with other parameters the same as the first case (black trace).

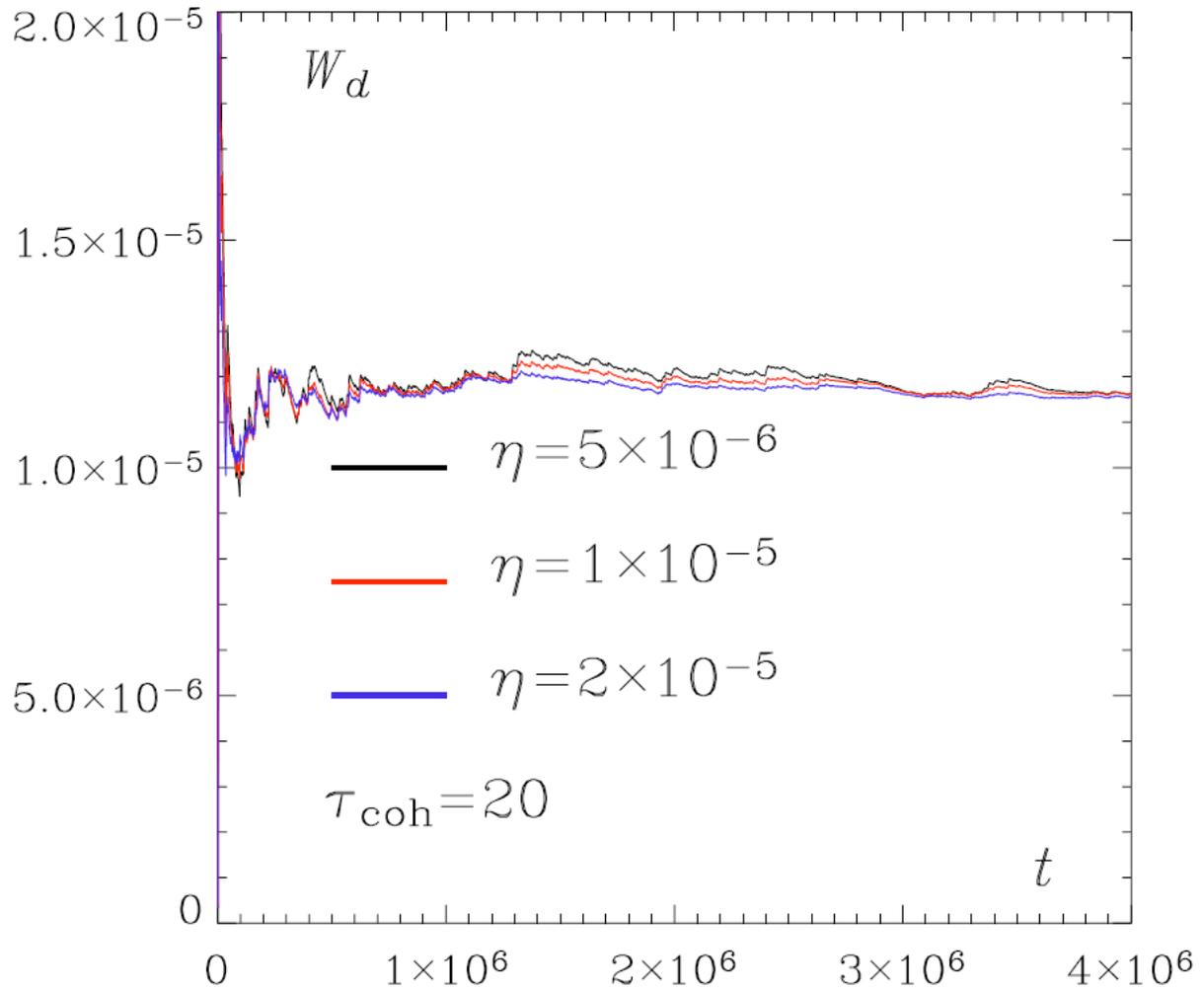

Figure 4. The time-averaged energy dissipation rate $\overline{W}_d$ as a function of time for the case with $\tau_{coh} = 20$, $\eta = 5 \times 10^{-6}$ (black trace). Also plotted are $\overline{W}_d$ for $\eta = 10^{-5}$ (red trace), and $\eta = 2 \times 10^{-5}$ (blue trace), with other parameters the same as the first case (black trace).

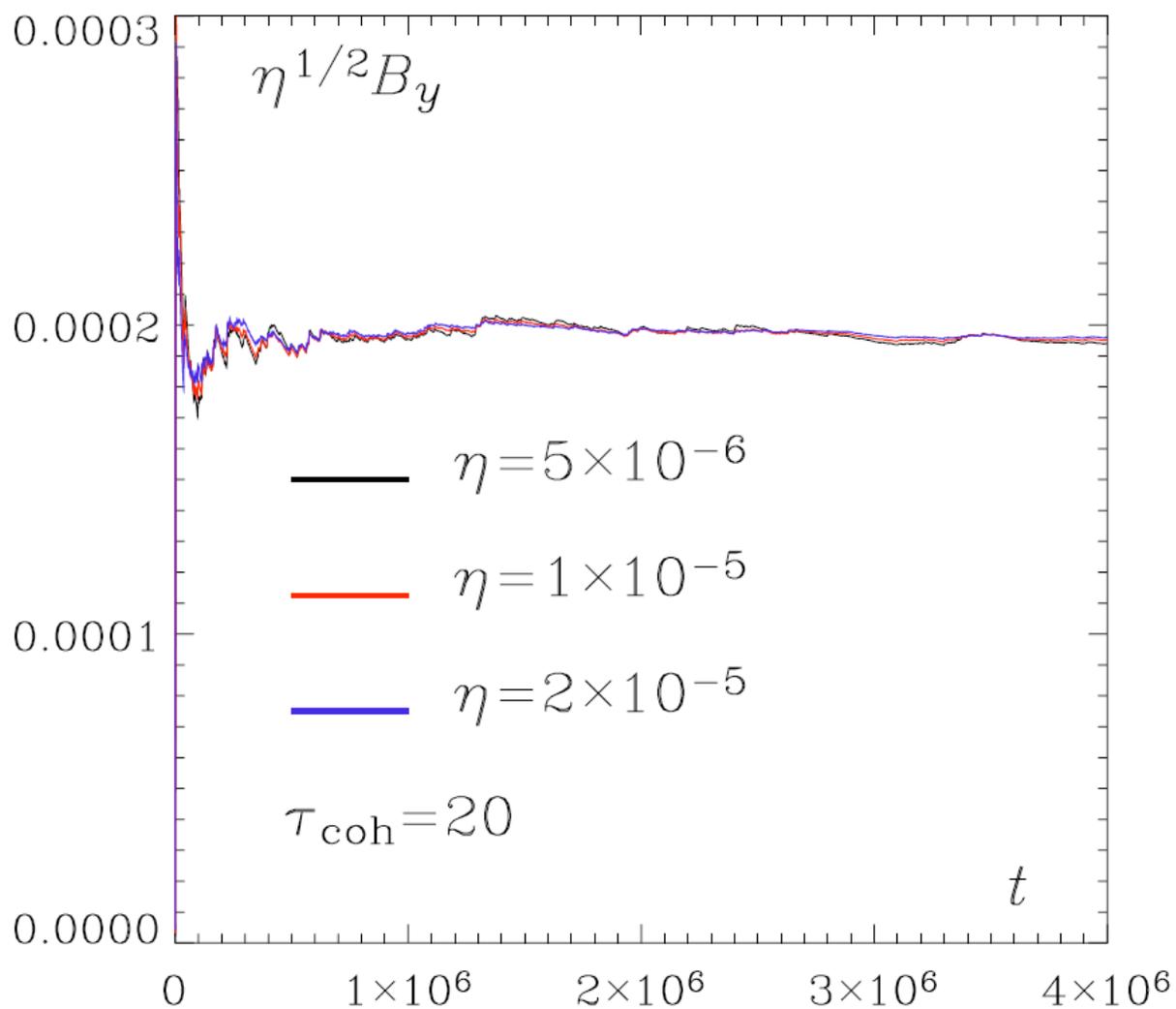

Figure 5. Root-mean-square $\overline{B}_y$ multiplied by $\eta^{1/2}$ as a function of time for the same set of runs as Figure 4.

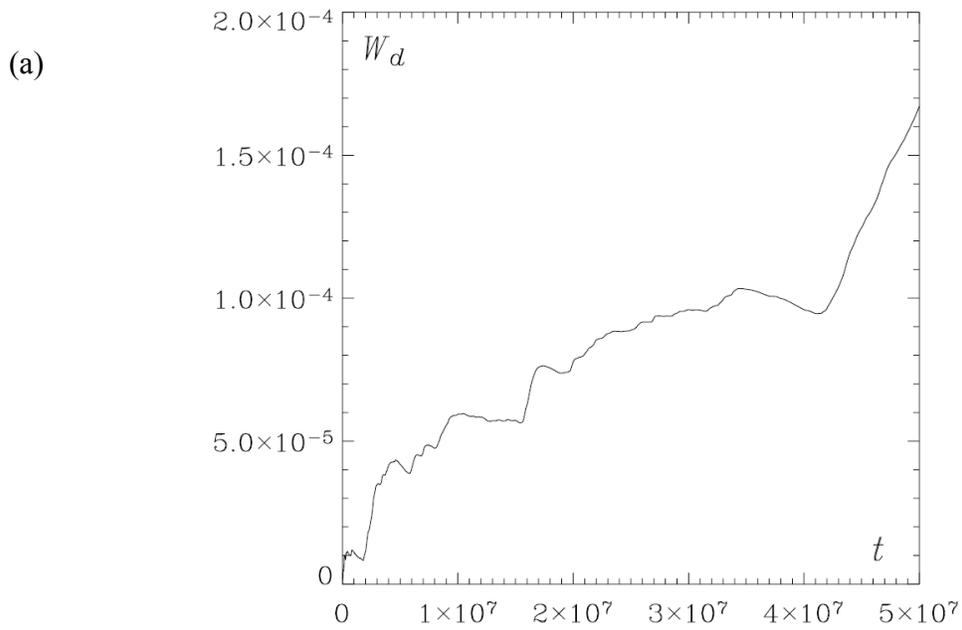

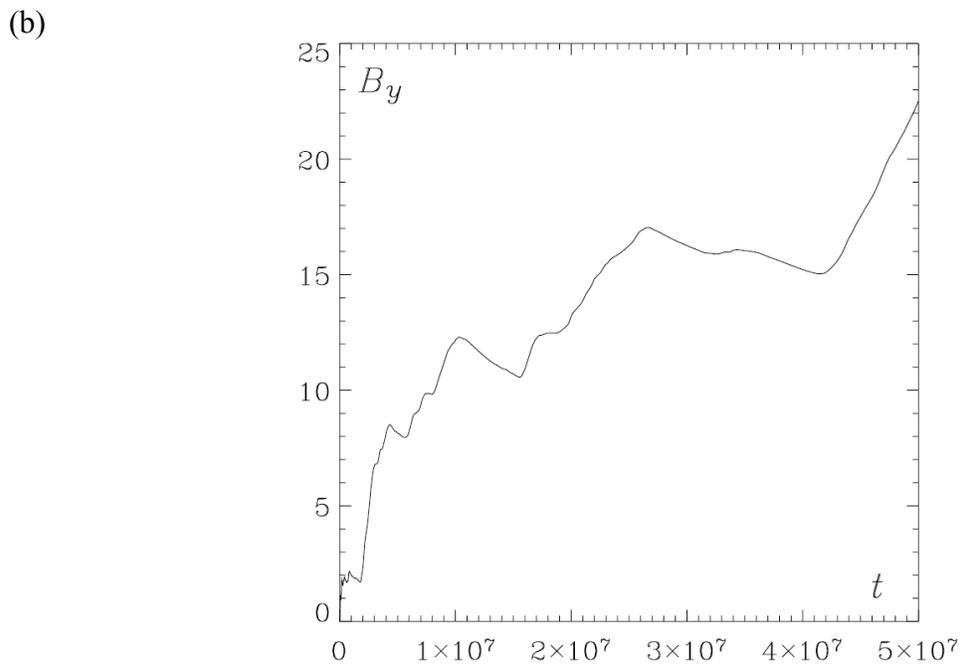

Figure 6. (a) Average energy dissipation rate $\overline{W}_d$ and (b) root-mean-square $\overline{B}_y$ as functions of time for the case with $\eta = 10^{-10}$, $\tau_{coh} = 600$ and other parameters the same as the run in Figure 5.

REFERENCES

Aschwanden, M. 2004, Physics of the Solar Corona (Berlin: Springer-Verlag)

Bhattacharjee, A. 2004, ARA&A, 42, 365

Close, R. M., Parnell, C. E., Longcope, D. W., & Priest, E. R. 2004, ApJ, 612, L81

Galsgaard, K., & Nordlund, A. 1996, JGR, 101, 13445

Gudiksen, B., & Nordlund, A. 2002, ApJ, 572, L113

———. 2005a, ApJ, 618, 1020

———. 2005b, ApJ, 618, 1031

Hagenaar, H. J. 2001, ApJ, 555, 448

Hendrix, D. L., Van Hoven, G., Mikic, Z., & Schnack, D. D. 1996, ApJ, 470, 1192

Klimchuk, J. A. 2006, Sol. Phys., 234, 41

Longcope, D. W., & Sudan, R. N. 1994, ApJ, 437, 491

Martin, S. F. 1988, Sol. Phys. 117, 243

Parker, E. N. 1972, ApJ, 174, 499

Parker, E. N. 1994, Spontaneous Current Sheets in Magnetic Fields: with Applications to Stellar X-Rays (New York: Oxford University Press)

Parnell, C. E. 2001, Sol. Phys., 200, 23

Priest, E. R., Heyvaerts, J. F., & Title, A. M. 2002, ApJ, 576, 533

Schrijver, C. J., et al. 1998, Nature, 394, 152

Schrijver, C. J., & Zwaan, C. 2000, Solar and Stellar Magnetic Activity, (Cambridge: Cambridge Univ. Press)

Title, A. M. 2000, Philos. Trans. R. Soc. London, A358, 657

van Ballegooijen, A. A. 1986, ApJ, 311, 1001